# Space-charge effects in low-energy flat-beam transforms

Scott B. Moroch and Timothy W. Koeth

*University of Maryland, College Park, Maryland 20742, USA*

Bruce E. Carlsten

*Los Alamos National Laboratory, Los Alamos, New Mexico 87545, USA*

Flat-beam transforms (FBTs) provide a technique for controlling the emittance partitioning between the beam's two transverse dimensions. To date, nearly all FBT studies have been in regimes where the beam's own space-charge effects can be ignored, such as in applications with high-brightness electron linacs where the transform occurs at high, relativistic, energies. Additionally, FBTs may provide a revolutionary path to high-power generation at high frequencies in vacuum electron devices where the beam emittance is currently becoming a limiting factor, which is the focus of this paper. Electron beams in vacuum electron devices operate both at a much lower energy and a much higher current than in accelerators and the beam's space-charge forces can no longer be ignored. Here we analyze the effects of space charge in FBTs and show there are both linear and nonlinear forces and effects. The linear effects can be compensated by retuning the FBT and by adding additional quadrupole elements. The nonlinear effects lead to an ultimate dilution of the lower recovered emittance and will lead to an eventual power limitation for high-frequency traveling-wave tubes and other vacuum electron devices.

## I. INTRODUCTION

Introduced about two decades ago, a flat-beam transform (FBT) [1] is a clever technique to "move" beam emittance from one transverse plane to the other while preserving the product of the emittances (and thus the beam's overall 4-dimensional phase-space area). A FBT works by applying an axial magnetic field at the location of the cathode which introduces an initial correlation between the two transverse dimensions. Three skew quadrupole magnets (meaning the quadrupoles are rotated 45° from their usual orientation) remove the correlations, leading to a beam with one final emittance lower than the beam's intrinsic emittance (i.e., the transverse emittance that would occur in the absence of the FBT) and the other final emittance larger than the beam's intrinsic emittance. Significantly, the product of the final emittances is equal to the square of the intrinsic emittance. FBTs have been demonstrated [2,3] and analyzed [4,5] and simple approaches to determining the skew quadrupole strengths have been published [6,7]. To date, space-charge effects in FBTs have been ignored with the exception of [5], where the space-charge effects at low energy, well before the skew quadrupole section at high energy, were studied. In this paper, alternatively, we consider the nature of the space-charge effects within the skew quadrupole section itself where the beam can be highly asymmetric.

The motivation for this work is the future application of FBTs to high-power high-frequency vacuum electron devices such as planar traveling-wave tubes (TWTs) [8,9]. Because of the decreasing beam tunnel area of cylindrical TWTs as their frequency increases, a planar geometry allows higher beam current and thus higher generated RF power. A FBT is especially well matched to a planar geometry, where the stable transport of a planar electron beam in a planar geometry requires a small emittance in the beam's narrow plane while a much higher emittance can be tolerated in the beam's wider plane. A typical beam voltage and current for such a device may be 20 kV and 5 A [8]. To emphasize the difference between this regime and that corresponding to a 20-A beam at 20 MeV from a photoinjector, such as used in [5], the 20-kV beam has *6 orders of magnitude higher beam perveance* than the 20-MeV beam. Thus, the space-charge effects in a FBT need to be re-examined.

The three major contributions of this paper are: (1) identification of the form of the linear space-charge effects in a FBT; (2) showing that the linear space-charge effects in a FBT can be completely removed with a combination of retuning the skew quadrupole strengths and by rotating the middle quadrupole so it adds some amount of a standard quadrupole field; and (3) evaluation of the residual emittance dilution due to the nonlinear space-charge effect from the nonuniformity of the beam density profile as the beam becomes pinched in one dimension within the skew quadrupole section.

Section II of this paper motivates this work, by describing how a FBT can be the basis for a revolutionary new architecture for high-frequency TWTs and other vacuum electron devices. We provide the necessary background for the following numerical and analytic studies in Section III. First, in Section III-A, we provide the basic formulas needed for designing a FBT. In Section III-B, we numerically extend symmetric-beam FBT theory to the case of an initially asymmetric beam. Specifically, we see that either the final lower or the final upper emittance can be increased, depending on the nature of how the initial beam emittances have been changed. The increase in the final lower emittance is the ultimate limitation of using a FBT to generate a high-power sheet beam for high-frequency RF generation.

In Section IV we describe our two modeling tools, PUSHER [6] and A Space Charge Tracking Algorithm (ASTRA) [10]. PUSHER only includes linear forces on the particles (it uses the Lawson linear space-charge approximation for an elliptical beam [11]) and is used to study the linear component of the beam's space-charge forces. ASTRA is fully nonlinear, with an advanced multipole space-charge routine fully capable of accurately modeling the nonlinear space-charge forces from an elliptical beam at any orientation. We provide baseline FBT simulation results for a beam without space charge for comparison with later results. In Section V, we calculate the linear space-charge forces of a rotated elliptical beam and show how some of these forces can be compensated by retuning the skew quadrupole strengths and also that the remaining forces can be compensated by simply rotating the middle skew quadrupole to introduce a component of a regular (non-skew) quadrupole field. In Section VI we evaluate the residual emittance dilution from the nonlinear component of the space-charge force for a 20-keV, 250-mA beam and extrapolate this result to estimate the maximum emittance improvement as a function of beam perveance.

## II. VACUUM ELECTRON DEVICE MOTIVATION

There is increasing interest to make high power (nominally kW to 10s of kW peak power) at high frequencies (nominally Ka-band to W-band, or roughly 30 GHz to 100 GHz), often with large bandwidths (10% or more), with compact vacuum electron devices (VEDs) [12-15] for radar, remote sensing, communications, and other applications. Reference [16] contains an excellent summary of high-frequency applications requiring higher power than currently available that are driving the development of new higher-frequency, higher-power sources and amplifiers.

Increasing the power of a VED can be achieved by increasing the beam current or by increasing the beam voltage, which makes the device less practical for many applications, plus adds extra length. The beam current can be increased by moving to a multi-beam device, as in a multi-beam klystron [17] or a multi-beam inductive-output tube (IOT) [18], or by employing a sheet electron beam [19-22]. While sheet beam devices have already been demonstrated with strong uniform solenoidal magnetic fields [21-25], periodic focusing [26-28] can lead to a more compact system without beam curling [29] and can be more desirable for many applications, especially those requiring high efficiency and/or compact sizes.

Sheet beams can be generated with initially round beams and focused into an elliptical shape with either magnetic quadrupole lenses [30,31] or an elliptical solenoid [31], or directly with an elliptically shaped electron gun [32,33]. (It is important here to point out that, in this paper, a "sheet" beam means a high-aspect ratio elliptical beam. A true rectangular sheet beam is actually harder to efficiently generate than a high-aspect ratio elliptical beam plus it introduces significant transport issues.) Traveling-wave tube slow-wave structures that are highly suited for sheet beams have been developed, including vane-loaded waveguides [34] and staggered vane-loaded waveguides [20,35].

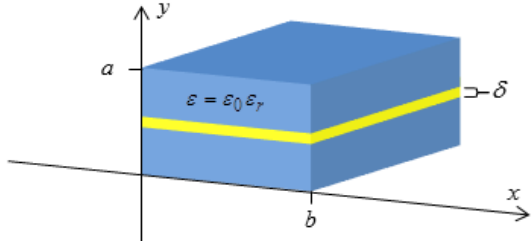

Fig. 1. Nominal TWT with a planar gap slow wave structure suitable for a high-power sheet electron beam. Two blocks of dielectric with constant $\varepsilon$ are separated by a vacuum gap $\delta$ shown in yellow. The external metallic boundary surrounding the structure is not shown.

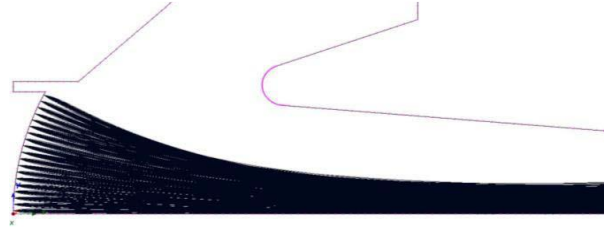

Fig. 2. Electron gun simulation showing the initial transverse beam velocity spread from the random thermal electronic motion at the cathode, leading to an initial beam emittance.

More recently, emerging low-loss, high-dielectric constant ceramics (like magnesium-calcium-titanate (MCT) from Trans-Tech, Inc [36] with a relative permittivity of 20 and a loss tangent of about $3 \times 10^{-4}$ at Ka-band) has motived the evaluation of axially uniform slow-wave structures [37,38,8,39,40]. While the NRL design [37] used a cylindrical geometry, the LANL design concept is planar (see Fig. 1), and suitable for a high-power electron beam.

One measure of an electron beam's quality is known as the normalized emittance, defined by

$$\varepsilon_{x,norm} = \gamma\beta\sqrt{\langle x^2\rangle\langle x'^2\rangle - \langle xx'\rangle^2} \quad (1)$$

for the horizontal dimension and in a similar manner for the vertical dimension, where the brackets indicate ensemble averages, the primes refer to axial derivatives, and $\gamma$ and $\beta$ are the beam's relativistic mass factor and velocity normalized to the speed of light $c$, respectively. A beam's emittance can be considered a pressure that defocuses the beam, similar to space charge. Beam emittance can be caused by, for example, the transverse thermal spread of the electron emission at the cathode (see Fig. 2) or by the electron flow around a cathode shadow grid. The thermal energy of the emission from the cathode provides a lower limit on the beam's transverse rms emittance, given by [41]

$$\varepsilon_{x,norm} = \sigma_{x,cath}\sqrt{\frac{kT_{cath}}{mc^2}} \quad (2)$$

where $\sigma_{x,cath} \equiv \sqrt{\langle x^2_{cath}\rangle}$ is the rms beam size at the cathode, $T_{cath}$ is the cathode temperature, $k$ is the Boltzmann constant, and $m$ is the mass of an electron. The emittance from a thermionic gun is typically about 1 $\mu$m for currents in the regime we're interested in (nominally 5 A at 20 kV, or 100 kW of peak beam power), i.e., a cathode using a long-life scandate cathode operating at 10 A/cm$^2$ [42] and an edge radius of 4 mm would then produce a 5-A electron beam with a thermal rms emittance of 0.9 $\mu$m at a cathode temperature of about 1200 K. Here, we will call the beam's initial thermal emittance plus any growth due to nonlinear fields in the gun diode region the beam's intrinsic emittance to differentiate it from emittance growths in later stages.

The beam emittance in the narrow plane (which we define here as $y$) drives the transport stability. First, the planar beam tunnel must be very narrow at high frequencies as the interaction field at the center of the tunner drops off as $cosh^{-1}(d\omega/2\beta c)$ [8], where $d$ is the tunnel width and $\omega$ is the RF frequency, which leads to a maximum practical rms beam size of about $\sigma_y \approx \beta c/4\omega$ where we are using $y$ for the narrow dimension and $x$ for the wide dimension. Assuming periodic focusing (with period $\lambda$) is used to confine the beam to this rms size with a magnetic field profile of $B = B_p\sin(\lambda z)$, the peak magnetic field needs to be

$$B_p = \left(\frac{cm}{e}\right)\sqrt{\frac{2}{\sigma_y}\left(\frac{I/I_A}{\beta\gamma}\frac{1}{\sigma_x+\sigma_y} + \frac{\varepsilon_{y,norm}^2}{\sigma_y^3}\right)} \quad (3)$$

where $m$ and $e$ are the electronic mass and charge, respectively, $I$ is the beam current, and $I_A$ is about 17 kA. Note as the beam size becomes smaller, the emittance term dominates the required focusing field.

Periodic focusing becomes unstable if the magnetic field strength is too high, with an absolute limit of

$$B_{p,limit} = 2\sqrt{1.3}\pi \frac{mc\beta\gamma}{e\lambda} \quad (4)$$

Ideally the peak focusing field is about a factor of 2 less than this limit. As there are physical limitations to how short the period $\lambda$ can be, Eqns. (3) and (4) tell us that the emittance must decrease as the frequency is increased to keep the beam transport stable. These arguments motivate the desire to use a FBT in the sheet-beam forming section of a high-frequency planar VED.

## III. BASIC CONCEPTS

Here we provide the key background material needed for the following FBT analysis in Sections IV-VI

### A. Basic eigen-emittance and FBT equations

A beam's eigen-emittances are conserved under symplectic transformations [4] (like those found in accelerator beam lines) and, in the absence of coupling between the transverse dimensions, the real transverse emittances are equal to the beam's eigen-emittances. If coupling is present, the real transverse emittances are larger than the eigen-emittances. Formulas for calculating the beam's eigen-emittances based on the beam's second moments are presented in the Appendix.

In a FBT, a nonzero axial magnetic field at the beam cathode can be used to decrease one transverse emittance at the expense of the other, while keeping the product constant. The configuration of a FBT is shown in Fig. 3.

In a FBT, the magnetic field on the cathode adds a canonical angular momentum to the beam, leading initially to transverse emittances of

$$\varepsilon_{trans} = \sqrt{\mathbf{L}^2 + \varepsilon_0^2} \quad (5)$$

where $\mathbf{L}$ is due to the angular momentum [4],

$$\mathbf{L} = \frac{e|B_{cath}|}{8\gamma\beta cm} R_{cath}^2 \quad (6)$$

in which $R_{cath}$ is the beam's radius at the cathode, $B_{cath}$ is the axial magnetic field at the location of the cathode, $e$ is the charge of an electron, and $\varepsilon_0$ is the intrinsic emittance of the gun (i.e., what the transverse emittances would be in the absence of the axial magnetic field on the cathode). Three skew quadrupoles are used to remove all *x-y* correlations (such that $\langle xy \rangle, \langle xy' \rangle, \langle x'y \rangle, \langle x'y' \rangle$ all vanish) and the final *x* and *y* emittances become these eigen-emittances (for the case where the angular momentum contribution is much larger than the intrinsic emittance) [4]

$$\varepsilon_{eig,-} = \frac{\varepsilon_0^2}{2\mathbf{L}} \quad (7)$$

$$\varepsilon_{eig,+} = 2\mathbf{L}. \quad (8)$$

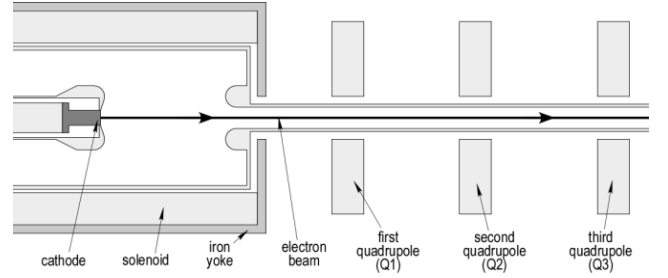

Fig. 3. Flat-beam transform configuration. A solenoid generates an axial magnetic field at the location of the cathode. Once the beam leaves the solenoidal field, it encounters three skew quadrupole which remove the *x-y* correlations. Image from [5].

Note that now one transverse emittance can be much larger than the other while keeping the product constant. To date, FBTs have only been demonstrated on RF photoinjectors, with a typical final emittance ratio $\varepsilon_{eig,+}/\varepsilon_{eig,-}$ of about 100. Reference [43] reports for the first time the development of a FBT for a low-voltage, high-frequency, sheet-beam traveling-wave tube, where the emittance in the beam's narrow dimension must be reduced for achieving stable transport and for ensuring decent interaction between the electron beam and the RF structure as described in the previous section. Because the eigen-emittances are recovered at low energy, the beam's space-charge forces change the skew quadrupole strengths for recovering the eigen-emittances and additionally lead to growth in the eigen-emittances.

A solution for the skew quadrupole strengths can be found in [6] and [7]. This solution is derived for a zero-emittance beam that is at a waist at the location of the first skew quadrupole by solving for the skew quadrupole strengths where all initial particle parameters lead to a final horizontal position and divergence of zero. This solution corresponds to the correct FBT solution for a nonzero emittance beam also at a waist at the location of the first quadrupole. Using

$$A = \frac{e}{2\gamma\beta mc} B_{cath} \frac{R_{cath}^2}{r^2} \quad (9)$$

to represent the angular rotation of the beam at the location of the first skew quadrupole and,

$$C_1 = \frac{e}{\gamma\beta mc} \int_{-\infty}^{\infty} B'_{Q1} \, dl \quad (10)$$

$$C_2 = \frac{e}{\gamma\beta mc} \int_{-\infty}^{\infty} B'_{Q2} \, dl \quad (11)$$

$$C_3 = \frac{e}{\gamma\beta mc} \int_{-\infty}^{\infty} B'_{Q3} \, dl \quad (12)$$

to represent the fields in the first, second, and third skew quadrupoles, respectively, and $D_{1-2}$ and $D_{2-3}$ to represent the distances between the centers of the first and second quadrupole and the second and third quadrupole, respectively, the skew quadrupole field strengths for the FBT are given by:

$$C_{2,0}^2 \frac{D_{2-3}}{D_{1-2}+D_{2-3}} - 2C_{2,0}A - \frac{1}{D_{1-2}D_{2-3}} = 0 \quad (13)$$

$$C_{1,0} = \frac{1}{2}\left(\frac{1}{C_{2,0}D_{1-2}D_{2-3}} + C_{2,0}\frac{D_{2-3}}{D_{1-2}+D_{2-3}}\right) \quad (14)$$

$$C_{3,0} = \frac{C_{2,0}\left(1 - \frac{D_{2-3}}{D_{1-2}+D_{2-3}}\right)}{1 - C_{2,0}^2 \frac{D_{2-3}}{D_{1-2}+D_{2-3}} D_{1-2}D_{2-3}} \quad (15)$$

where the "0" subscripts represent solutions for zero current. Note that there are two solutions for $C_2$ which then also lead to two different solutions for $C_1$ and $C_3$. However, one solution (the one with the "+" sign in front of the square root in the quadratic solution) leads to much smaller quadrupole field strengths which is preferable. With $A > 0$ and $C_2 > 0$, the smaller eigen-emittance ends up in the horizontal dimension.

## B. Extension of the FBT to an initially asymmetric beam

We anticipate nonlinear space-charge forces in the FBT will degrade the eigen-emittances along the lines Sun predicted it would for photoinjectors [5]. The mechanism examined in [5] was the nonlinear growth of beam divergence due to a nonuniform beam density distribution in the first cell of an RF photoinjector which contributes to a growth in the beam's emittance to be added in quadrature to the beam's intrinsic emittance $\varepsilon_0$ in Eqns. (5) and (7). In Sun's case, the high beam energy at the location of the transform itself ensured that space-charge forces in the skew quadrupole section could be ignored.

Here, however, the beam still has low energy when it travels through the skew quadrupole section and we anticipate the beam will experience a similar nonlinear eigen-emittance growth due to the fact it is compressed into a high aspect ratio ellipse. We expect the main effect to be a nonlinear increase in the divergence in the narrow dimension.

To provide some initial insights, the plots in Fig. 4 show what happens to the beam's eigen-emittances for a beam initially with 1 μm intrinsic emittances with a 1-mm radius at the cathode immersed in a 500-G axial magnetic field as the emittance of one of the dimensions is increased. We see the growth is predominantly in the lower eigen-emittance if the growth is due to an increase in beam divergence and predominantly in the upper eigen-emittance if the growth is due to an increase in the beam size. While there are significant differences for the beam within the skew quadrupole section (e.g., the beam transverse correlations are being modified by the skew quadrupoles), this suggests the nonlinear space-charge force will largely degrade the lower eigen-emittance which will lead to an ultimate limitation in the usefulness of a FBT for a low-energy beam.

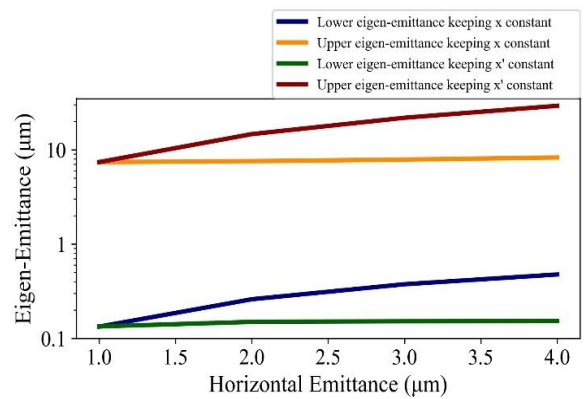

Fig. 4. Increase in eigen-emittances as one of the initial transverse emittances is increased by up to a factor of four. If the increase is due to an increase in divergence, the lower eigen-emittance is increased.

## IV. MODELING TOOLS

We used the codes PUSHER [6] and ASTRA [10] to model the FBT. PUSHER is a simple time-stepping, particle pushing code following an axial slice of the beam and including only linear space-charge forces. We typically ran it with 5000 particles to represent the beam and it would model the FBT in about two seconds, making it useful for scoping studies and to clarify the effects of the linear components of the space charge. To include the nonlinear space-charge effects in the beam transformer, we used ASTRA. ASTRA includes a full 3D space charge algorithm, in which the bunch is modeled as a distribution of macro-charges [44]. The bunch is broken into Cartesian cells of constant charge density and Poisson's equation is solved in the rest frame of the bunch, using a Fast-Fourier Transform method. A python-based wrapper was used in conjunction with ASTRA to optimize the skew quadrupole strengths to recover the eigen-emittances.

We first verified the codes with the beam transformer theory by studying the zero-current case. For the simulations in this manuscript, we consider a 1-mm radius, 20-keV electron beam with a normalized rms emittance of 1 $\mu$m, produced from a cathode submersed in a solenoidal field. The solenoid field has a peak strength of 500 G and induces an angular momentum contribution to the emittance of $\mathbf{L} = 3.66$ $\mu$m. This yields theoretical, rms emittances of $\varepsilon_y = 7.46$ $\mu$m and $\varepsilon_x = 0.133$ $\mu$m at the end of the transformer. According to Eqn. (7), this corresponds to a final emittance ratio of 53.77. These parameters are summarized in Table 1.

| Beam energy | 2- keV |
| --- | --- |
| Beam current | 0 to 250 mA |
| Radius at cathode | 1 mm |
| Magnetic field at cathode | 500 G |
| Intrinsic emittance | 1.00 |
| Angular momentum term | 3.66 $\mu$m |
| Lower eigen-emittance | 0.133 $\mu$m |
| Upper eigen-emittance | 7.32 $\mu$m |
| Emittance ratio | 53.8 $\mu$m |

Table 1. Nominal beam parameters for the FBT.

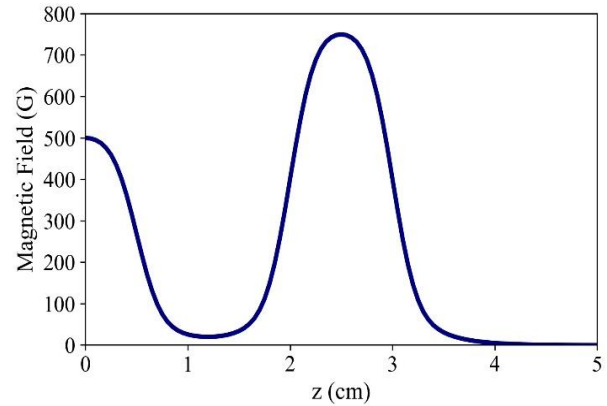

fig. 5. Plot of the axial magnetic field profile in cathode region of the ASTRA simulation

There were some minor differences in how we set up the PUSHER and ASTRA simulations. In PUSHER we were able to initiate the particles with the proper amount of angular momentum and immediately inject them into the first skew quadrupole. In ASTRA, we needed to model the magnetic field at the location of the cathode, which extended longitudinally enough to impact the beam's injection into the first skew quadrupole, located at $z = 5$ $cm$. As a result, we added a second, focusing, solenoid centered at $z = 2.5$, as seen in Fig. 5, to bring the beam to a waist at the position of the first quadrupole. The solenoidal field on axis $(x = y = 0)$ was modeled in POISSON SUPERFISH [45] and imported into ASTRA which computed the off-axis field components to third order.

Since the waist radius is slightly different than the injection radius (and the radius used in the PUSHER simulations), the skew-quadrupole strengths needed to recover the eigen-emittances are slightly different. To show this, the variation in skew-quadrupole strengths as a function of beam size in the first skew quadrupole is shown in Table 2. Since PUSHER and ASTRA had somewhat different injection conditions, the optimized skew-quadrupole settings differed and, as a result, here we focus more on the physics than the precise values of the settings.

| $r_{waist}$ (mm) | $B_{Q1}$ (T/m) | $B_{Q2}$ (T/m) | $B_{Q3}$ (T/m) |
| --- | --- | --- | --- |
| 0.8 | -7.93 | 0.236 | -0.119 |
| 0.9 | -6.32 | 0.298 | -0.151 |
| 1.0 | -5.18 | 0.365 | -0.186 |
| 1.1 | -4.352 | 0.439 | -0.225 |
| 1.2 | -3.735 | 0.517 | -0.268 |

Table 2. Skew quadrupole settings as a function of the injector beam radius.

The ASTRA results of modeling the zero current FBT are shown in Fig. 6, demonstrating the recovery of the eigen-emittances, as predicted. The variation in the eigen-emittances in the cathode region are attributed to the solenoid fringe fields changing the beam's angular momentum. The python wrapper computed the beam emittance ratio for over 3000 different skew quadrupole settings. This data is depicted in Fig. 7. Note there appears to be a "trough"

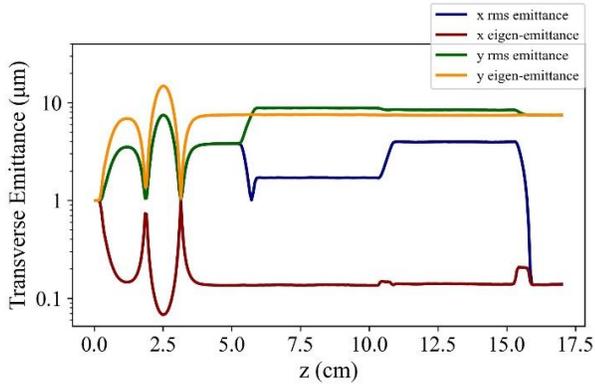

Fig. 6. Evolution of the beam emittances and eigen-emittances in the nominal flat beam transformer with zero current using the parameters in Table. 1. The changes in the eigen-emittances in the cathode region may be attributed to the non-symplectic nature of the solenoid fringe fields.

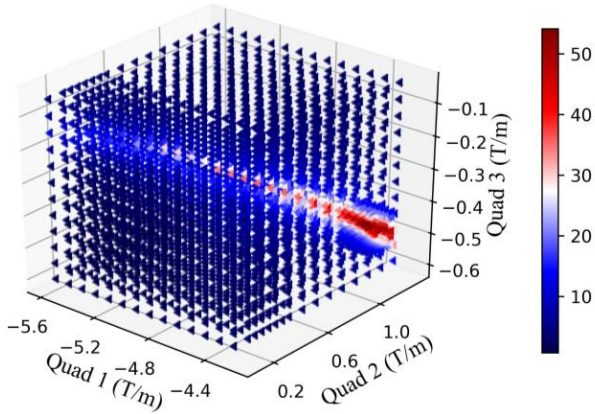

Fig 7. 3D plot of emittance ratio as a function of skew quadrupole parameters

of skew quadrupole settings with higher emittance ratios.

## V. EFFECT OF LINEAR SPACE CHARGE

Next, we consider an elliptical beam with some tilt, as shown in Fig. 8, which is representative of the beam within the FBT section.

### A. Linear space-charge theory

A uniform filled ellipse has space charge forces,

$$F_{\tilde{x},sc} = \frac{eI_b}{4\pi\varepsilon_0 \beta_b \gamma_b^2 c \sigma_{\tilde{x}}(\sigma_{\tilde{x}}+\sigma_{\tilde{y}})} \tilde{x} \quad (16)$$

$$F_{\tilde{y},sc} = \frac{eI_b}{4\pi\varepsilon_0 \beta_b \gamma_b^2 c \sigma_{\tilde{y}}(\sigma_{\tilde{x}}+\sigma_{\tilde{y}})} \tilde{y} \quad (17)$$

in the rotated coordinate frame $(\tilde{x}, \tilde{y})$ where the beam is an upright ellipse [11] (Fig. 8) and where $\sigma_{\tilde{x}}$ and $\sigma_{\tilde{y}}$ are the rms horizontal and vertical beam sizes in that frame. The rotation angle $\theta$ in terms of the second moments of the $(x, y)$ coordinate system is given by

$$\theta = \frac{1}{2} tan^{-1}\left(\frac{2\langle xy \rangle}{\sigma_x - \sigma_y}\right) \quad (18)$$

The space-charge forces in the $(x, y)$ coordinate system are then

$$F_{x,sc} = xk\left(\frac{cos^2\theta}{\sigma_{\tilde{x}}} + \frac{sin^2\theta}{\sigma_{\tilde{y}}}\right) + yk\frac{sin2\theta}{2}\left(\frac{1}{\sigma_{\tilde{x}}} - \frac{1}{\sigma_{\tilde{y}}}\right) \quad (19)$$

$$F_{y,sc} = yk\left(\frac{cos^2\theta}{\sigma_{\tilde{y}}} + \frac{sin^2\theta}{\sigma_{\tilde{x}}}\right) + xk\frac{sin2\theta}{2}\left(\frac{1}{\sigma_{\tilde{x}}} - \frac{1}{\sigma_{\tilde{y}}}\right) \quad (20)$$

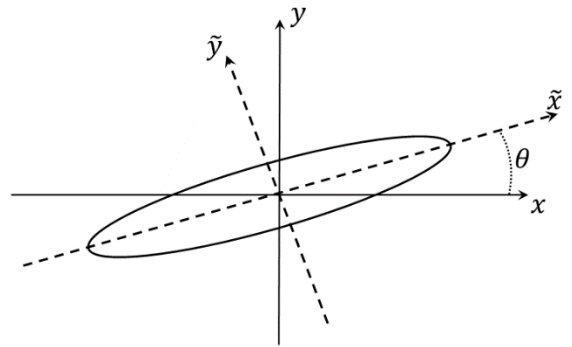

Fig. 8. Geometry used in the text. The beam is an upright ellipse in a new coordinate system $(\tilde{x}, \tilde{y})$ rotated an angle $\theta$ from the $(x, y)$ coordinate system.

where $k = \frac{eI_b}{4\pi\varepsilon_0 \beta_b \gamma_b^2 c(\sigma_{\tilde{x}}+\sigma_{\tilde{y}})}$. Importantly, the second, correlated terms, are exactly of the form of the force from a skew quadrupole, and, as such, can be compensated by a change in the skew quadrupole strengths. Note the coefficients of the first, non-coupling, term are a mixture of a symmetric focusing (or defocusing) force and an asymmetric quadrupole force (focusing in one plane and defocusing in the other). Also note that the forces coupling $x$ and $y$ cannot have a second part equivalent to the symmetric focusing (or defocusing) term because that would add angular momentum to the beam, which the beam's self-force cannot do.

Let us now consider the effect of the non-coupling space-charge forces with a simplifying calculation. We assume the beam vector is $(x_0, x'_0, y_0, x'_0)$ just before the second skew quadrupole and that we can lump the integrated effect of the non-coupling terms just after the second skew quadrupole as $(x', y') = (x'_0 + C_2 y_0 + \Gamma x_0, y'_0 + C_2 x_0 \pm \Gamma y_0)$, where the plus sign is for a focusing (defocusing) contribution and the minus sign represents a quadrupole field. We also assume that the values of the skew quadrupoles $(C_1, C_2, C_3) = (C_{1,0}, C_{2,0}, C_{3,0})$ represent the solution shown in Eqns. (13-15) for the case there is no space charge (i.e., the non-coupling force $\Gamma$ vanishes). We assume the strength of the first skew quadrupole is unchanged but allow the strengths of the second the third skew quadrupoles to change as $(C_1, C_2, C_3) = (C_{1,0}, C_{2,0} + \tilde{C}_2, C_{3,0} + \tilde{C}_3)$. Using the fact that the final horizontal position and divergence $(x_f, x'_f)$ both vanish for the solution where $\Gamma = 0$, we can find $(x'_0, y'_0)$ in terms of $x_0$ and $y_0$ ($x'_0 = x_0/D_{2-3} - C_{2,0}y_0$ and $y'_0 = x_0(-C_{2,0} - 1/C_{3,0}D_{2-3}^2) - y_0 D_{2-3}$) and we find

$$x_f = \Gamma D_{2-3} x_0 + \tilde{C}_2 D_{2-3} y_0 \qquad (21)$$

$$x'_f = \left(\Gamma + \tilde{C}_2 C_3 D_{2-3} - \frac{\tilde{C}_3}{C_{3,0}D_{2-3}}\right)x_0 + \qquad (22)$$
$$(\tilde{C}_2 \pm \Gamma C_3 D_{2-3})y_0$$

The final $x$ emittance will vanish if the ratio $x_f/x'_f$ is independent of $x_0$ and $y_0$. Setting the $x_0$ and $y_0$ coefficients to have the same ratio, we find this solution for $\tilde{C}_2$:

$$\tilde{C}_2 = \frac{\tilde{C}_3}{2C_3 C_{3,0}D_{2-3}^2} \pm \sqrt{\left(\frac{\tilde{C}_3}{2C_3 C_{3,0}D_{2-3}^2}\right)^2 \pm \Gamma^2} \qquad (23)$$

This solution suggests three important properties: (1) there is an infinite number of zero-emittance solutions even in the case the non-coupling forces vanish, $\Gamma = 0$; (2) there is always a well-behaved solution for the case of the plus sign in front of the $\Gamma^2$ term, which corresponds to the focusing (defocusing) non-coupling term; and (3) there may not be a solution for the case of the minus sign in front of the $\Gamma^2$ term (since the term with $\tilde{C}_3$ cannot be made arbitrarily large). Note that the first property explains the trough of solutions shown in Fig. 7. The second and third properties suggest that the focusing (defocusing) component of the space charge force can be compensated by a change in the skew quadrupole settings but the asymmetric non-coupling component of the space-charge force cannot. Thus we anticipate that a regular quadrupole field will need to be added to fully recover the eigen-emittances. Also, importantly, this result implies we can add focusing as needed within the FBT to ensure good beam transport without affecting the ability to recover the eigen-emittances.

**B. Linear space-charge simulations**

We verified these observations using PUSHER, with simulation results shown Figs. 9-11. The blue curve in Fig. 9 corresponds to a 50-mA beam current where an additional regular quadrupole field is superimposed with the field from the second skew quadrupole. The strength of the regular quadrupole field is varied and plotted as a fraction of the skew quadrupole field strength (which was optimized at each fraction value). The best achievable recovered eigen-emittances only with skew quadrupoles (i.e., a fraction of zero) is 0.1413 $\mu$m.

The red curve corresponds to the zero-current case where a symmetric focusing field is superimposed with the field from the second skew quadrupole, also shown as a function of its fractional strength as compared to the skew quadrupole field (which also varied by about 10% over the range of the plot while being about 30% lower than for the 50-mA case). These plots confirm the suggestions from the simple analysis: the skew quadrupole strengths can be re-tuned to eliminate the coupling term from the linear space charge; a regular quadrupole field can be used to eliminate the non-coupling term; and a symmetric focusing field doesn't substantially change the performance of the FBT. (We should point out the lower eigen-emittance for the zero-current case was numerically found to be about 0.133 $\mu$m, just as predicted, but the lower eigen-emittance of the 50-mA case was about 0.135 $\mu$m. We attribute this difference to numerical noise in the code's space-charge model leading to some variation in the beam's total angular momentum.

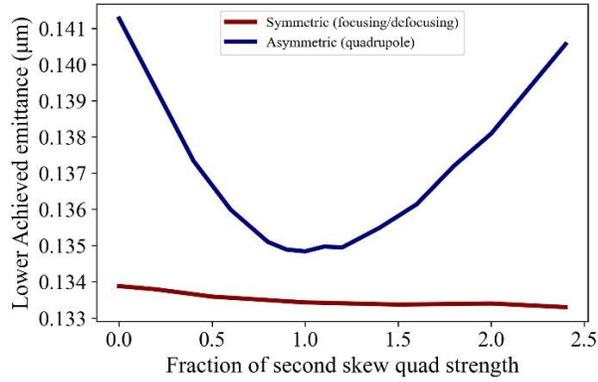

Fig. 9. PUSHER simulations demonstrating the ability to eliminate the emittance growth due to the non-coupling quadrupole field due to space charge with a regular quadrupole field superimposed with that of the second skew quadrupole (blue) and the relatively minor effect of a focusing field superimposed with the second skew quadrupole field, for zero beam current (red).

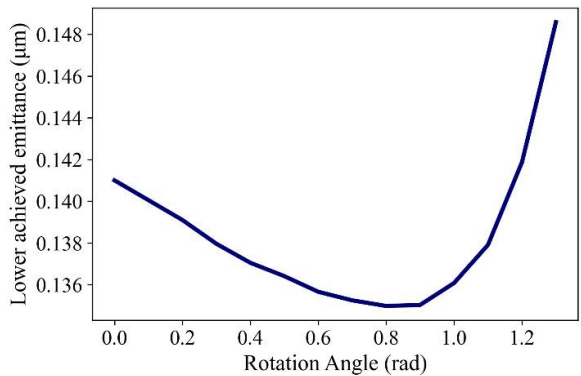

Fig. 10. Compensation of the non-coupling quadrupole field due to space charge as a function of the rotation angle of the second skew quadrupole. The strength of the quadrupole was adjusted to maintain about the same amount of skew quadrupole variation as in Fig. 9.

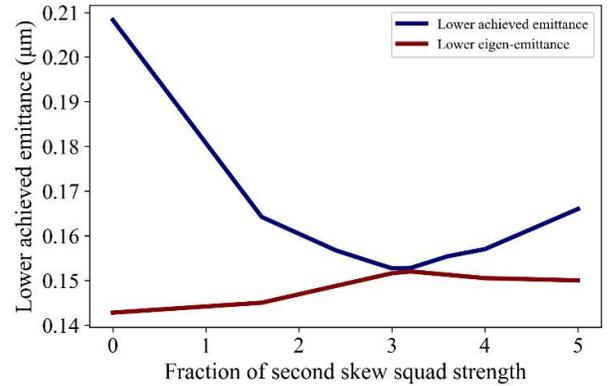

Fig. 11. The blue curve shows that the lower achieved emittance as a function of the applied quadrupole field at the location of the second skew quadrupole, as a fraction of the skew quadrupole strength. The red curve shows the value of the lower eigen-emittance for that case. The strength of all three skew quadrupoles were varied for all fractions to minimize the lower achieved emittance.

A superimposed skew quadrupole field and regular quadrupole field can be generated by simply rotating the skew quadrupole, with the equivalent results as the blue curve in Fig. 9 shown in Fig. 10. The same compensation for a 250-mA beam is shown by the blue curve in Fig. 11. To show the lower eigen-emittance is essentially recovered, we plot the calculated lower eigen-emittance in red as a function of the additional quadrupole field (as it also varies due to numerical noise in the simulation). Considering the difference of the red and blue curves allows us to see that the linear components of the space-charge force is in fact removed with proper tuning of the additional regular quadrupole field.

## VI. EFFECT OF NON-LINEAR SPACE CHARGE

In this section, we consider the effects of non-linear space charge forces, with a specific focus on the skew-quadrupole region where the beam is highly asymmetric.

### A. Non-linear space-charge simulations

To model the effects of non-linear space charge forces in a FBT, a 0.11 ns bunch, consisting of $10^5$ electrons, was tracked in ASTRA. To emulate a DC beam (as present in VEDs) the emittances and transverse charge density profiles were calculated only for a central slice (5%) of the bunch. Therefore, despite the presence of longitudinal space charge effects in the overall simulated bunch, they are suppressed in the central slice. In the cathode region, ASTRA's cylindrically symmetric space-charge solver is employed and the bunch is split into radial and longitudinal cells. In the FBT section, the beam becomes highly asymmetric and requires ASTRA's 3D Cartesian-based algorithm. The bunch is split into a $32 \times 32 \times 32$ Cartesian grid, corresponding to approximately 3 macro-charges per cell for computing the space-charge fields. As described in Section IV for the zero-current case, a python-based wrapper was employed to vary both the three skew quadrupole strengths and rotation angles, to recover the eigen-emittances. Contrary to the previous simulations, the non-linear forces lead to strong coupling between the quadrupole strengths and the beam eigen-emittances, making it substantially more difficult to find a solution. The optimized solution for a 250-mA beam current is shown in Fig. 12 and demonstrates the ability to recover the eigen-emittances with proper tuning of the quadrupoles. The emittance growth is dominated in the skew-quadrupole region, with the lower (0.25 $\mu$m) and upper (8.62 $\mu$m) emittances growing by 60% and ~13%, respectfully, compared to the zero-current case.

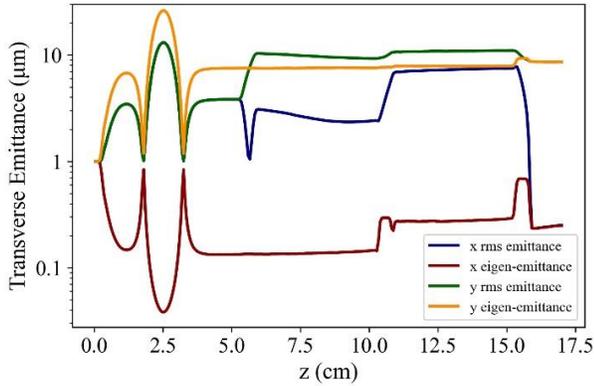

Fig. 12. Optimized solution for the 250-mA case in ASTRA, demonstrating successful recovery of beam eigen-emittances.

This emittance growth may be attributed to the nature of the FBT, in which the narrow plane of the beam is squeezed, yielding a Gaussian charge-density profile. The transverse beam profile in the rotated-frame of the beam, mid-way between the second and third quadrupoles is plotted in the top of Fig. 13. It is evident that the charge density is non-linear in the narrow-dimension ($x$) and nearly uniform in the wider-dimension ($y$). As a verification, the bunch was binned in the narrow dimension, revealing a Gaussian profile in the bottom of Fig. 13.

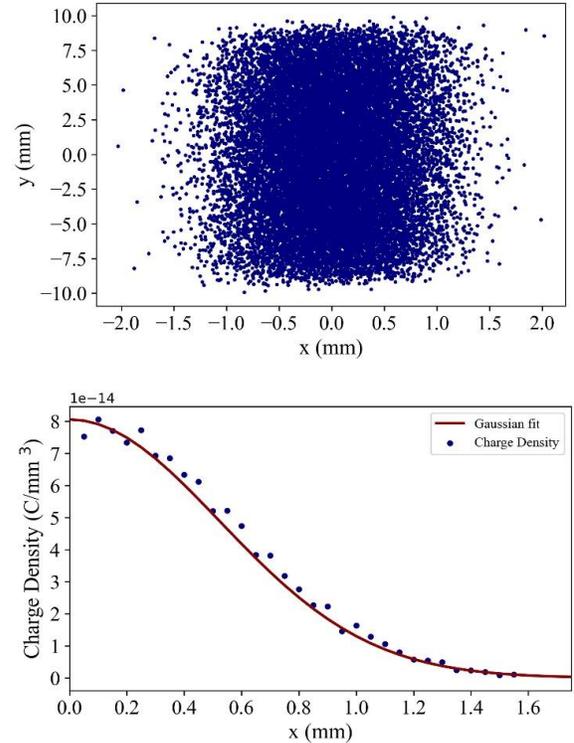

Fig. 13. (Top) Beam cross section in the beam-rotated frame at 12.26 cm, showing a highly asymmetric beam which has nearly uniform density in $y$. (Bottom) The density in $x$ very nearly follows a Gaussian shape, as predicted by [46].

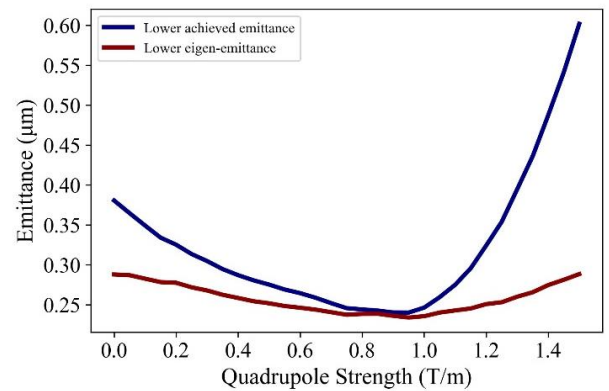

Fig. 14. Lower-eigen emittance as a function of the strength of a superimposed, regular-quadrupole at the location of the second skew-quadrupole.

In analogy with Fig. 9, Fig. 14 demonstrates the ability to recover the lower eigen-emittance by introducing a regular quadrupole field, superimposed over the second skew quad. Without this quadrupole field, the best attainable lower emittance was 0.38 $\mu$m, supporting the physics described in Section V-A.

## B. Estimate of the effect from nonlinear space charge

As described, we attribute the growth in the lower eigen-emittance to the nonlinear space-charge forces, which result from the beam's nonuniform density in its narrow dimension, as it gets squeezed in the FBT. For comparison, the rms beam sizes in the rotated frame of the beam (as in Fig. 8), are plotted in Fig. 15. To verify this emittance growth, we proceed by formulating an analytical expression for the emittance growth of a beam with nonlinear charge density in the narrow-dimension ($x$) and uniform density in $y$ and $z$-dimensions.

Following [5], we find that the change in momentum in the squished dimension is given by:

$$\Delta p_x = \frac{1}{mc} \int_0^{t_0} F_x \, dt \approx \frac{1}{mc^2} \int \frac{F_x}{\beta} \, dz \quad (24)$$

The charge density of the beam in this dimension can be written as:

$$\rho(x) = \rho_0 e^{-\frac{x^2}{2\sigma_x^2}} \approx \rho_0 \left(1 - \frac{1}{2\sigma_x^2} x^2 + \frac{1}{4\sigma_x^4} x^4 \ldots\right) \quad (25)$$

and applying Gauss' law to the geometry in Fig. 16, we find

$$E_x = \frac{1}{\varepsilon_0} \int_0^x \rho(x') \, dx' \quad (26)$$

$$E_x = \frac{\rho_0}{\varepsilon_0} \int_0^x e^{-\frac{x'^2}{2\sigma_x^2}} \, dx' \quad (27)$$

$$E_x \approx \frac{\rho_0 x}{\varepsilon_0} \left[1 - \frac{1}{6\sigma_x^2} x^2\right] \quad (28)$$

with a net horizontal force of:

$$F_x = \frac{e\rho_0 x}{\gamma^2 \varepsilon_0} \left[1 - \frac{1}{6\sigma_x^2} x^2\right] \quad (29)$$

The increase in horizontal momentum is then:

$$\Delta p_x = \frac{e}{mc^2} \frac{\rho_0 x}{\varepsilon_0} \left[1 - \frac{1}{6\sigma_x^2} x^2\right] \int_0^{z_0} \frac{1}{\beta \gamma^2} \, dz \quad (30)$$

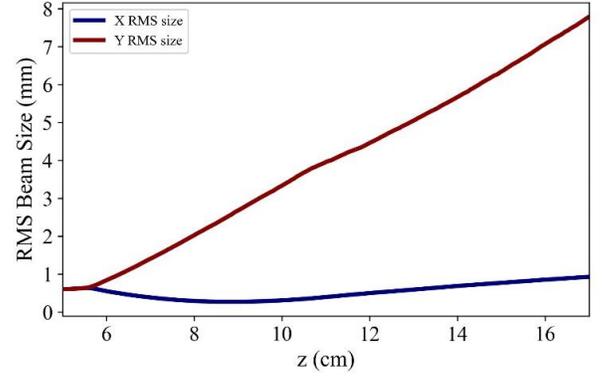

Fig. 15. Comparison of the larger and smaller beam sizes through the FBT in the beam-rotated frame.

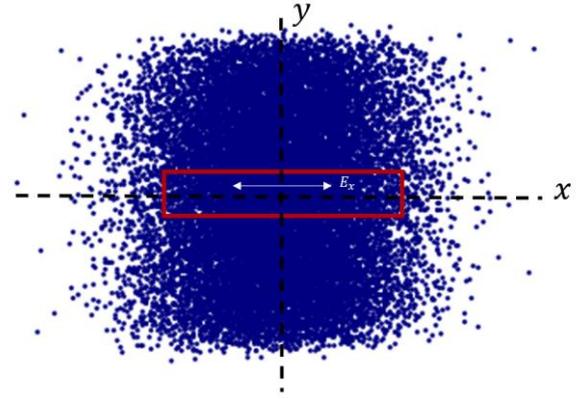

Fig. 16. Geometry used to calculate the $x$-components of the electric field, as a function of $x$.

and since there is no energy gain,

$$\Delta p_x = \frac{\rho_0 z_0 e}{\beta \gamma^2 mc^2 \varepsilon_0} \left[x - \frac{1}{6\sigma_x^2} x^3\right] \quad (31)$$

To calculate the increase in emittance, we need:

$$\langle \Delta p_x^2 \rangle = \frac{\int_{-\infty}^{\infty} \Delta p_x^2 \rho(x) \, dx}{\int_{-\infty}^{\infty} \rho(x) \, dx} = \left[\frac{\rho_0 z_0 e}{\beta \gamma^2 mc^2 \varepsilon_0}\right]^2 \left[\frac{15}{36} \sigma_x^2\right] \quad (32)$$

$$\langle x^2 \rangle = \frac{\int_{-\infty}^{\infty} x^2 \rho(x) \, dx}{\int_{-\infty}^{\infty} \rho(x) \, dx} = \sigma_x^2 \quad (33)$$

$$\langle x \Delta p_x \rangle = \frac{\int_{-\infty}^{\infty} x \Delta p_x \rho(x) \, dx}{\int_{-\infty}^{\infty} \rho(x) \, dx} = \frac{\rho_0 z_0 e \sigma_x^2}{2\beta \gamma^2 mc^2 \varepsilon_0} \quad (34)$$

The corresponding emittance growth is:

$$\varepsilon_{sc} = \sqrt{\langle \Delta p_x^2 \rangle \langle x^2 \rangle - \langle x \Delta p_x \rangle^2} = \frac{\sqrt{6}}{6} \left[\frac{\rho_0 z_0 e}{\beta \gamma^2 mc^2 \varepsilon_0}\right] \sigma_x^2 \quad (35)$$

Or, after converting the charge density to the current using $I = v\rho_0 \sigma_x \Delta y \sqrt{2\pi}$,

$$\varepsilon_{sc} = \frac{\sqrt{3\pi}}{6} \left[\frac{1}{\beta^2 \gamma^2}\right] \frac{z_0 \sigma_x}{\Delta y} \frac{I}{I_A} \quad (36)$$

where $\Delta y$ is the vertical height of the beam. We can now predict the emittance growth from this effect. Looking at Fig. 13, we can estimate this growth over a length, $z_0$, of about 10 cm with the beam size of $\Delta y \approx 18$ mm and $\sigma_x \approx 0.5$ mm, at $z = 12$ cm. For a 250-mA beam current, this emittance growth is about 0.22 $\mu$m, corresponding to a final lower eigen-emittance of 0.26 $\mu$m, when added in quadrature with the predicted eigen-emittance of 0.136 $\mu$m for the zero-current case. This is in good agreement with the actual simulated lower eigen-emittance of 0.25 $\mu$m (Fig. 12).

### C. Impact of nonlinear space-charge on using a FBT to reduce the emittance in a single dimension

We can extend the above analysis to make a general statement regarding the maximum beam current for a flat beam transformer. For the flat beam transformer to be useful we need

$$\varepsilon_- = \sqrt{\left(\frac{\varepsilon_0^2}{2L}\right)^2 + \varepsilon_{sc}^2} < \varepsilon_0^2 \qquad (37)$$

or

$$\varepsilon_{sc} < \varepsilon_0 \qquad (38)$$

since we want $\frac{\varepsilon_0^2}{2L}$ to be small, which leads to a limit of about 1 A. It is important to note that this value may be increased if the beam narrow dimension can be kept larger on average with symmetric focusing elements which would not affect our ability to achieve the lower eigen-emittance as shown by Fig. 9.

### VII. CONCLUSIONS

In this paper, we reviewed FBTs, and extended the existing theory to include space-charge dominated beams. Our motivation for this was to determine if FBTs could be applied to future low-voltage vacuum electron devices that require high-brightness sheet beams. We first analyzed the effects of the linear space-charge components which prevent a standard FBT from completely recovering the beam eigen-emittances. We uncovered, analytically, the ability to compensate the emittance growth introduced by these linear space-charge forces in a FBT by re-tuning the skew-quadrupole strengths and introducing a regular, superimposed quadrupole field. Simulations, conducted with a particle-pushing code (PUSHER), supported these results, and demonstrated the ability to fully recover the beam eigen-emittances. To study the effects of nonlinear space-charge, we modeled the beamline in ASTRA, which includes a full 3D space-charge algorithm. These simulations indicated that in the skew-quadrupole region, the charge-density profile in the narrow-plane of the beam became Gaussian, leading to nonlinear space-charge forces and a growth of the lower eigen-emittance of about ~60% for our nominal parameters. We conclude that this emittance growth, caused by squeezing of the beam, represents a fundamental limiting factor of FBTs and will need to be carefully considered when approaching higher power and higher frequency sheet-beam traveling wave tubes. It may be possible to mitigate this nonlinear effect by minimizing the squeezing by adding additional quadrupole focusing.

### APPENDIX

In this appendix, we provide an explicit eigen-emittance calculation. This development follows [47] and [48] based on the background in [4]. We start with the beam second moment matrix,

$$\sigma = \begin{pmatrix} \langle x^2 \rangle & \langle xx' \rangle & \langle xy \rangle & \langle xy' \rangle \\ \langle xx' \rangle & \langle x'^2 \rangle & \langle x'y \rangle & \langle x'y' \rangle \\ \langle xy \rangle & \langle x'y \rangle & \langle y^2 \rangle & \langle yy' \rangle \\ \langle xy' \rangle & \langle x'y' \rangle & \langle yy' \rangle & \langle y'^2 \rangle \end{pmatrix}$$

We assume there is a transfer matrix $R$ that is symplectic ($R^T J R = J$) and that can diagonalize the beam matrix,

$$\sigma_{diag} = R\sigma R^T = \begin{pmatrix} \varepsilon_+ & 0 & 0 & 0 \\ 0 & \varepsilon_+ & 0 & 0 \\ 0 & 0 & \varepsilon_- & 0 \\ 0 & 0 & 0 & \varepsilon_- \end{pmatrix}$$

where $\varepsilon_+ > \varepsilon_-$ are the eigen-emittances and $J$ is the four-dimensional skew symmetric matrix

$$J = \begin{pmatrix} 0 & 1 & 0 & 0 \\ -1 & 0 & 0 & 0 \\ 0 & 0 & 0 & 1 \\ 0 & 0 & -1 & 0 \end{pmatrix}.$$

Two conservation properties (conservation of the determinant and of the trace $-\frac{1}{2}Tr((\sigma J)^2)$ [49]) allow us to find the eigen-emittances. First, from

$$(\sigma_{diag}J)^2 = \begin{pmatrix} -\varepsilon_+^2 & 0 & 0 & 0 \\ 0 & -\varepsilon_+^2 & 0 & 0 \\ 0 & 0 & -\varepsilon_-^2 & 0 \\ 0 & 0 & 0 & -\varepsilon_-^2 \end{pmatrix}$$

we have (using $\sigma_{diag} = R\sigma R^T$, $R^T J R = J$, and the cyclic property of traces $Tr(ABC) = Tr(BCA) = Tr(CAB)$)

$$Tr\left((\sigma_{diag}J)^2\right) = Tr((J\sigma)^2) = Tr((\sigma J)^2)$$
$$= -2(\varepsilon_+^2 + \varepsilon_-^2)$$

and from the determinant we have

$det(\sigma_{diag}) = det(R)det(\sigma)det(R^T) = det(\sigma) = \varepsilon_+^2 \varepsilon_-^2$ .

Combining the trace and determinant we find

$$Tr^2((\sigma J)^2) - 16 det(\sigma) = 4(\varepsilon_+^4 - 2\varepsilon_+^2\varepsilon_-^2 + \varepsilon_-^4)$$
$$= \left(2(\varepsilon_+^2 - \varepsilon_-^2)\right)^2$$

resulting in

$$\varepsilon_\pm = \frac{1}{2}\left(-tr((\sigma J)^2) \pm \sqrt{tr^2((J\sigma)^2) - 16 det(\sigma)}\right)^{1/2}.$$

Numerically, this can be simplified further to facilitate the eigen-emittance calculations. First, we transform the beam matrix $\sigma$ with this transfer matrix

$$R_0 = \begin{pmatrix} 1 & 0 & 0 & 0 \\ -\frac{\langle xx'\rangle}{\langle x^2\rangle} & 1 & 0 & 0 \\ 0 & 0 & 1 & 0 \\ 0 & 0 & -\frac{\langle yy'\rangle}{\langle y^2\rangle} & 1 \end{pmatrix}$$

which simplifies the beam matrix to the form

$$\sigma_0 = R_0 \sigma R_0^T = \begin{pmatrix} \sigma_{0,1}^2 & 0 & D & B \\ 0 & \sigma_{0,2}^2 & E & F \\ D & E & \sigma_{0,3}^2 & 0 \\ B & F & 0 & \sigma_{0,4}^2 \end{pmatrix}.$$

The eigen-emittances are then

$$\varepsilon_\pm^2 = U \pm V$$

where,

$$U = \frac{1}{2}\left(\sigma_{0,1}^2\sigma_{0,2}^2 + \sigma_{0,3}^2\sigma_{0,4}^2 - 2BE + 2FD\right)$$

and

$$V^2 = \frac{1}{4}\left(\sigma_{0,1}^2\sigma_{0,2}^2 + \sigma_{0,3}^2\sigma_{0,4}^2 - 2BE + 2FD\right)^2 - \left(\sigma_{0,1}^2\sigma_{0,2}^2\sigma_{0,3}^2\sigma_{0,4}^2 - F^2\sigma_{0,1}^2\sigma_{0,3}^2 - E^2\sigma_{0,1}^2\sigma_{0,4}^2 - D^2\sigma_{0,2}^2\sigma_{0,4}^2 - B^2\sigma_{0,2}^2\sigma_{0,3}^2 - 2EBDF + D^2F^2 + E^2B^2\right).$$

---